# Nanosecond electro-optic switching of a liquid crystal

Volodymyr Borshch, Sergij V. Shiyanovskii, and Oleg D. Lavrentovich*

Chemical Physics Interdisciplinary Program, Liquid Crystal Institute,
Kent State University, Kent, OH, 44242, USA.
*Correspondence to: olavrent@kent.edu

**Abstract**: Electrically induced reorientation of nematic liquid crystal (NLC) molecules caused by dielectric anisotropy of the material is a fundamental phenomenon widely used in modern technologies. Its Achilles heel is a slow (millisecond) relaxation from the field-on to the field-off state. We present an electro-optic effect in an NLC with a response time of about 30 ns to both the field-on and field-off switching. This effect is caused by the electric field induced modification of the order parameters (EMOP) and does not require reorientation of the optic axis (director).

Nematic liquid crystals (NLCs) have revolutionized the way optical information is presented and processed [1]. The main feature that makes NLCs unique as an electro-optic medium is their long-range orientational order. NLC molecules are of an anisometric shape, in most cases resembling an elongated rod. Their average orientation is called the director $\hat{\mathbf{n}}$. The director is also the optic axis of an NLC with the birefringence $\Delta n = n_e - n_o$, where $n_e$ and $n_o$ are the extraordinary and ordinary refractive indices, respectively. Electro-optical applications exploit the so-called Frederiks effect, i.e., the reorientation of $\hat{\mathbf{n}}$ in a low-frequency electric field caused by anisotropy of dielectric susceptibility $\Delta\varepsilon = \varepsilon_\parallel - \varepsilon_\perp > 0$; $\varepsilon_\parallel$ and $\varepsilon_\perp$ are the permittivities measured parallel to $\hat{\mathbf{n}}$ and perpendicular to it, respectively. When field $\mathbf{E}$ is applied, the director $\hat{\mathbf{n}}$ realigns along $\mathbf{E}$ if $\Delta\varepsilon > 0$ and perpendicular to $\mathbf{E}$ if $\Delta\varepsilon < 0$. The reorientation time is approximately $\tau_{on}^F \approx \gamma / \varepsilon_0 |\Delta\varepsilon| E^2$, where $\gamma$ is the rotational viscosity and $\varepsilon_0$ is the electric constant. A strong field can realign an NLC rather quickly, within 100 ns [2]. However, a slow relaxation to the field-off state creates a bottleneck. When the field is switched off, the elastic nature of NLC forces $\hat{\mathbf{n}}$ to return to its original orientation set up by the surface treatment of the cell's plates (surface anchoring). The relaxation time of this passive process is determined by the elastic constant $K$ of the NLC and by the thickness $d$ of the cell, $\tau_{off}^F \approx \gamma d^2 / K\pi^2$. For typical $K = 10$ pN, $d = 5\,\mu\text{m}$, $\gamma = 0.1\,\text{Pa}\cdot\text{s}$, the relaxation is slow, $\tau_{off}^F \approx 25\,\text{ms}$.

In this work, we demonstrate an electro-optic effect in which both the field-on and field-off switching is fast, on the order of nanoseconds and tens of nanoseconds. The effect is based on electrically induced modification of the order parameters (EMOP) of NLCs rather than on the Frederiks reorientation of $\hat{\mathbf{n}}$.

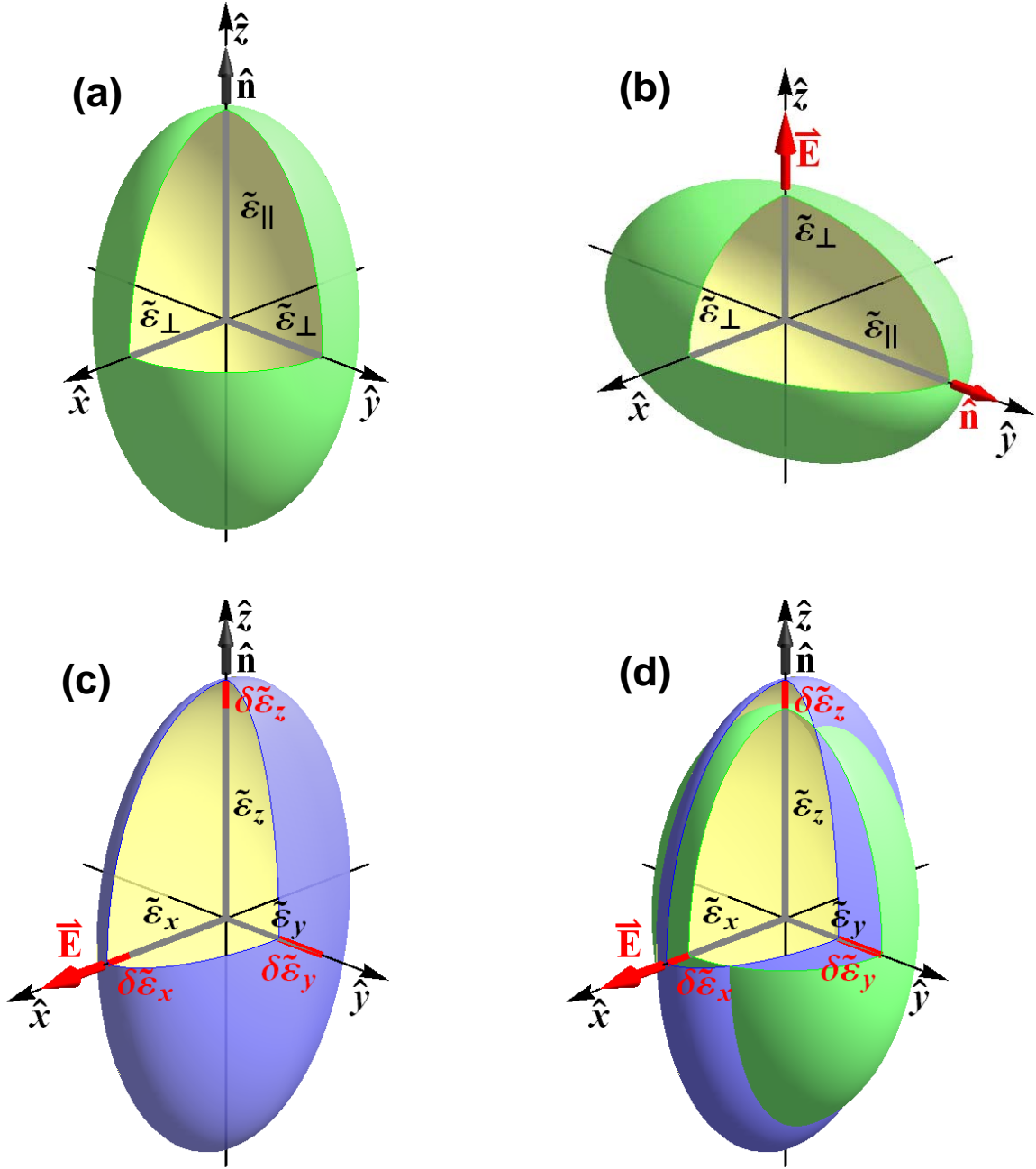

**Figure 1.** Ellipsoid of optic tensor. (**a**) Uniaxial NLC with $\Delta n > 0$, $\Delta\varepsilon < 0$, and $\hat{\mathbf{n}}$ along the $z$-axis. (**b**) Frederiks transition: $\hat{\mathbf{n}}$ reoriented by electric field $\mathbf{E}$ applied along the $z$-direction. (**c**) Electrically modified order parameter (EMOP) effect: Electric field $\mathbf{E}$ along the $x$-axis increases $\tilde{\varepsilon}_x$ and $\tilde{\varepsilon}_z$, while decreasing $\tilde{\varepsilon}_y$. (**d**) Field-off and field-on optical ellipsoids are shown together for comparison.

The difference between the two types of electro-optic response is schematically illustrated in Fig.1, using the concept of the dielectric tensor at optical frequencies (optic tensor). The two principal components of the optic tensor are $\tilde{\varepsilon}_\| = n_e^2$ and $\tilde{\varepsilon}_\perp = n_o^2$, such that $\Delta\tilde{\varepsilon} = \tilde{\varepsilon}_\| - \tilde{\varepsilon}_\perp > 0$, Fig. 1a. Consider the case when the low-frequency dielectric anisotropy is negative, $\Delta\varepsilon = \varepsilon_\| - \varepsilon_\perp < 0$. In the Frederiks effect, $\mathbf{E}$ is applied parallel to $\hat{\mathbf{n}}$, so that the director reorients to become perpendicular to $\mathbf{E}$, Fig.1b [1, 3]. In the EMOP effect, $\mathbf{E}$ is applied perpendicularly to $\hat{\mathbf{n}}$, to avoid director reorientation. The electric field changes the components of the optic tensor, Fig.1c: $\tilde{\varepsilon}_x = \tilde{\varepsilon}_\perp + \delta\tilde{\varepsilon}_x$; $\tilde{\varepsilon}_y = \tilde{\varepsilon}_\perp + \delta\tilde{\varepsilon}_y$; $\tilde{\varepsilon}_z = \tilde{\varepsilon}_\| + \delta\tilde{\varepsilon}_z$. These changes should manifest itself in phase retardation of light travelling through the NLC and thus provide the means for electro-optical switching. For example, when the NLC is viewed between two crossed polarizers, the change in phase retardation is converted into a change in light intensity, similar to the Frederiks effect. The principal difference is that the EMOP is a microscopic effect; its response time is determined by the coupling of the optic tensor of an NLC to $\mathbf{E}$. Such a response should be significantly faster than the off time $\tau_{off}^F \propto d^2$ of the Frederiks effect determined by macroscopic effects. Fig. 2a demonstrates the main result of this work, an electro-optic switching of the NLC with the response time well below 100 ns *to both the field on and field off* driving.

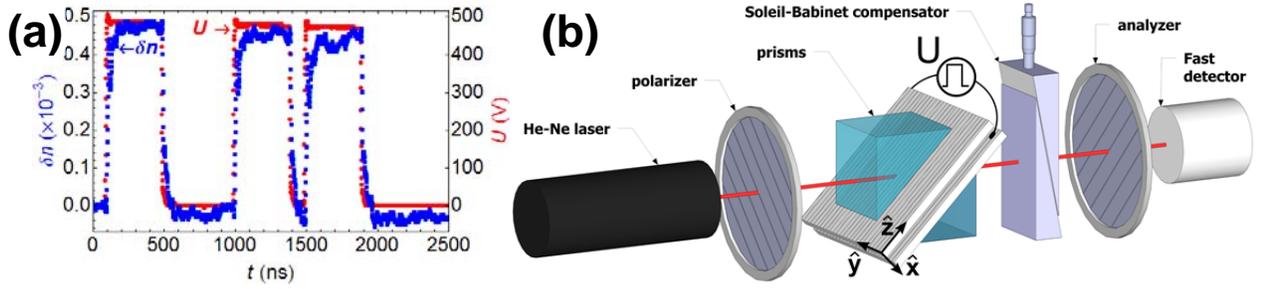

Figure 2. (**a**) EMOP effect: field-induced changes in birefringence $\delta n$ (blue) follow both the "on" and "off" edges of voltage pulses (red) with a characteristic response time of about 30 ns; NLC CCN-47; $T = 45$ °C. (**b**) Experimental setup: NLC cell sandwiched between two 45-degree prisms.

The very possibility of the field-induced modification of the nematic order parameter (OP) is recognized and well known [4-7]. For example, if **E** is parallel to $\hat{\mathbf{n}}$, it causes a uniaxial modification of $\Delta \varepsilon = \varepsilon_\parallel - \varepsilon_\perp$ and $\Delta \tilde{\varepsilon} = \tilde{\varepsilon}_\parallel - \tilde{\varepsilon}_\perp$, see e.g. Ref.[5], where the field-induced changes are described by the modification of the orientational ordering of the long molecular axes. The field applied normally to $\hat{\mathbf{n}}$ can lift the degeneracy of the transversal permittivity, producing a biaxial modification [5, 6, 8-11]. The field can also modify the fluctuations of $\hat{\mathbf{n}}$ [4, 11-15]. The effect of fluctuations is detrimental; as when the field is switched off, they relax to their field-free spectrum slowly, with a characteristic time up to $\tau_{off}^F$. We avoid the contribution of director fluctuations to the optical response by a special design of light propagation, as explained below.

We used 4'-butyl-4-heptyl-bicyclohexyl-4-carbonitrile (CCN-47) (Nematel GmbH, Germany) with the NLC phase in the temperature range $T = 31\text{-}58$ °C. CCN-47 is dielectrically negative, $\Delta \varepsilon < 0$. At $T = 40$ °C, $\Delta n = n_e - n_o = 0.029$ and $\Delta \varepsilon = -5.1$ at field frequencies (1-50) kHz. The NLC cell is 4.2 μm thick. The anchoring direction is along the $z$-axis in the plane of the cell, Fig.2b. The field is applied across the cell, along the $x$-axis, by two indium tin oxide (ITO) electrodes of a small area, $A = 9 \text{ mm}^2$. Voltage pulses of amplitude $U_0$ up to 1 kV with nanosecond rise and fall fronts were generated by a pulse generator HV 1000 (Direct Energy). Their profile was experimentally determined with an oscilloscope Tektronix TDS 2014 (sampling rate 1GSa/s). The field changes the optic tensor:

$$\delta\tilde{\varepsilon}_x = -\tfrac{1}{3}\delta\tilde{\varepsilon}_u + \tfrac{1}{2}\delta\tilde{\varepsilon}_b + \left(\langle n_x^2 \rangle_E - \langle n_x^2 \rangle_0\right)\left(n_e^2 - n_o^2\right),$$

$$\delta\tilde{\varepsilon}_y = -\tfrac{1}{3}\delta\tilde{\varepsilon}_u - \tfrac{1}{2}\delta\tilde{\varepsilon}_b, \qquad (1)$$

$$\delta\tilde{\varepsilon}_z = \tfrac{2}{3}\delta\tilde{\varepsilon}_u - \left(\langle n_x^2 \rangle_E - \langle n_x^2 \rangle_0\right)\left(n_e^2 - n_o^2\right).$$

Here $\delta\tilde{\varepsilon}_u$ is the field-induced enhancement of optical anisotropy caused by the uniaxial OP change; $\delta\tilde{\varepsilon}_b$ is the field-induced biaxiality of the optic tensor; $\langle n_x^2 \rangle$ is the average of the fluctuating director components along the $x$-axis that depends on the electric field; the subscripts "E" and "0" indicate whether the value corresponds to the field-on or field-off case respectively. We use a probing light beam that impinges on the NLC cell at 45°; the light is polarized at 45° to the plane of incidence, Fig.2b. The field-induced quantities $\delta\tilde{\varepsilon}_x$, $\delta\tilde{\varepsilon}_y$, and $\delta\tilde{\varepsilon}_z$ are small; thus the field-induced effective birefringence can be expanded in the power series, with the linear term corresponding to our set-up:

$$\delta n = \frac{n_o^2\left(2n_o^2 - n_p^2\right)\delta\tilde{\varepsilon}_z + n_p^2 n_e^2 \delta\tilde{\varepsilon}_x - 2n_e n_o^3 \delta\tilde{\varepsilon}_y}{4 n_e n_o^3 \sqrt{n_o^2 - n_p^2/2}}. \qquad (2)$$

Because the refractive indices of CCN-47, $n_o = 1.47$, $n_e = 1.50$, and the prisms, $n_p = 1.52$, are close, the contribution of slowly relaxing director fluctuations is practically eliminated, and the field-induced change $\delta n$ of birefringence is determined by $\delta\tilde{\varepsilon}_u$ and $\delta\tilde{\varepsilon}_b$:

$$\delta n = \frac{1}{2\sqrt{n_o^2 + n_e^2}} \left( \delta\tilde{\varepsilon}_u + \frac{3}{2}\delta\tilde{\varepsilon}_b \right). \tag{3}$$

Dynamics of $\delta n(t)$ in response to the voltage pulses was determined by measuring the intensity of a He-Ne laser beam (wavelength $\lambda = 633$ nm) probing the cell as shown in Fig.2b, $I(t) = I_0 \sin^2\left\{ \frac{\pi\left[\delta n(t) + \Delta n_{eff}\right]d}{\lambda} + \frac{\phi_{SB}}{2} \right\}$, where $\Delta n_{eff}$ is the effective birefringence without the field; $\phi_{SB}$ is the adjustable phase of the Soleil-Babinet compensator. The phase $\phi_{SB}$ was adjusted in order to (a) maximize the sensitivity of measurements to $\delta n(t)$ by setting $\phi_{SB}$ to one of the $\phi_m = \pi\left(m + \frac{1}{2}\right) - \frac{2\pi\Delta n_{eff} d}{\lambda}$, where $m$ is an integer, and (b) verify that the intensity change $\Delta I$ is caused by the EMOP effect ($\Delta I \to -\Delta I$ when $\phi_m \to \phi_{m+1}$) and not by the light scattering ($\Delta I \to \Delta I$ when $\phi_m \to \phi_{m+1}$) [16].

The most striking feature of the EMOP effect in Fig.2a is that both the field-on and field-off switching show extremely fast (tens of nanoseconds) response, a phenomenon very different from the Frederiks transition. In what follows, we analyze the EMOP dynamics and demonstrate that (a) the field-on and field-off response times are practically the same, and (b) the field-induced birefringence is a sum of the uniaxial $\delta\tilde{\varepsilon}_u$ and biaxial $\delta\tilde{\varepsilon}_b$ contributions to the optic tensor changes.

Figure 3a shows the birefringence change caused by a single voltage pulse of a duration 394 ns. To analyze the dependence $\delta n(t)$, we model the dynamics of the field-induced uniaxial $\delta\tilde{\varepsilon}_u$ and biaxial $\delta\tilde{\varepsilon}_b$ contributions to the optic tensor by the Landau-Khalatnikov approach [17]:

$$\tau_i \frac{d\delta\tilde{\varepsilon}_i(t)}{dt} = \alpha_i E^2(t) - \delta\tilde{\varepsilon}_i(t), \tag{4}$$

where the subscript "$i$" should read "$u$" if the equation describes the uniaxial contribution and "$b$" for the biaxial effect; $\tau_u$ and $\tau_b$ are the corresponding relaxation times; $\alpha_u$ and $\alpha_b$ are the susceptibilities to the applied field. Equation (4) has a general solution

$$\delta\tilde{\varepsilon}_i(t) = \int_0^t \frac{\alpha_i E^2(t')}{\tau_i} \exp\left(\frac{t'-t}{\tau_i}\right) dt' \tag{5}$$

that is used to fit the experimental $\delta n(t)$. In the integrand, the time dependence $E(t)$ of the field acting on the NLC cell is described by a sum of exponential functions

$$E(t_{on} \leq t \leq t_{off}) = \frac{U_0}{d}\left[\frac{e^{-(t-t_{on})/\tau_a}}{1-\tau_{RC}/\tau_a} - \frac{e^{-(t-t_{on})/\tau_{on}}}{1-\tau_{RC}/\tau_{on}} - \frac{(\tau_a - \tau_{on})e^{-(t-t_{on})/\tau_{RC}}}{(\tau_a - \tau_{RC})(1-\tau_{on}/\tau_{RC})}\right]$$

$$E(t > t_{off}) = E(t_{off})e^{-(t-t_{off})/\tau_{RC}} + \frac{U_0}{d(1-\tau_{RC}/\tau_{off})}\left[e^{-(t-t_{off})/\tau_{off}} - e^{-(t-t_{off})/\tau_{RC}}\right], \tag{6}$$

where $\tau_{RC} = RC \approx 7$ ns is the $RC$-time of the cell, $R \approx 43\ \Omega$ is the resistance of cell electrodes, $C \approx 160$ pF is the cell capacitance measured at $T = 45$ °C. The characteristic times of the voltage pulse, namely, the rise time of the front edge $\tau_{on}$, the fall down time of the rear edge $\tau_{off}$, and the time of slow decay of the pulse amplitude $\tau_a$, were obtained by fitting the pulse profile, Fig.3:

$$U(t < t_{on}) = 0;\ U(t_{on} \leq t \leq t_{off}) = U_0\left[e^{-(t-t_{on})/\tau_a} - e^{-(t-t_{on})/\tau_{on}}\right],\ \text{and}\ U(t > t_{off}) = U(t_{off})e^{-(t-t_{off})/\tau_{off}},$$

$$\tag{7}$$

where $t_{on}$ and $t_{off}$ are the moments when the field is switched on and off, respectively. With the known $E(t)$, we fit the experimental $\delta n(t)$ in Fig.3a using the solutions (5) for the uniaxial and biaxial modifications of the optic tensor. The fitting parameters are $\tau_u$, $\tau_b$, $\alpha_u$, and $\alpha_b$.

The fitting shows that the response $\delta n(t)$ to the voltage change (including the "on" and "off" parts) is well described by the sum of two exponential processes of approximately equal amplitudes but with distinct relaxation times. One process is "slow", with the relaxation time ~30 ns; the second process (dashed line in Fig.3a) is very fast, with a relaxation time ~2 ns or even less (at the limit of our experiment accuracy).

The temperature dependences of $\alpha_u$ and $\alpha_b$ as one approaches the nematic-to-isotropic phase transition are very different from each other, Fig.3c. The value of $\alpha_u$ increases dramatically, while $\alpha_b$ shows a small decline. This distinct behavior suggests that the slower process characterized by $\alpha_u$ should be attributed to the dynamics of uniaxial contribution $\delta\tilde{\varepsilon}_u$ and that the faster process with $\alpha_b$ should be attributed to the biaxial contribution $\delta\tilde{\varepsilon}_b$. The reason is that the free energy density near the nematic-to-isotropic transition has a vanishing dependence on the uniaxial OP and a sharp parabolic dependence on the biaxial OP[4]. The increase of susceptibility $\alpha_b$ at lower temperatures can be attributed to proximity of a hidden uniaxial-to-biaxial nematic phase transition. This transition is not observed experimentally, as the material transforms into a smectic A phase upon cooling. The latter might also explain why $\alpha_u$ increases at lower temperatures: formation of fluctuative smectic clusters near the nematic-to-smectic transition can be enhanced in presence of the electric field.

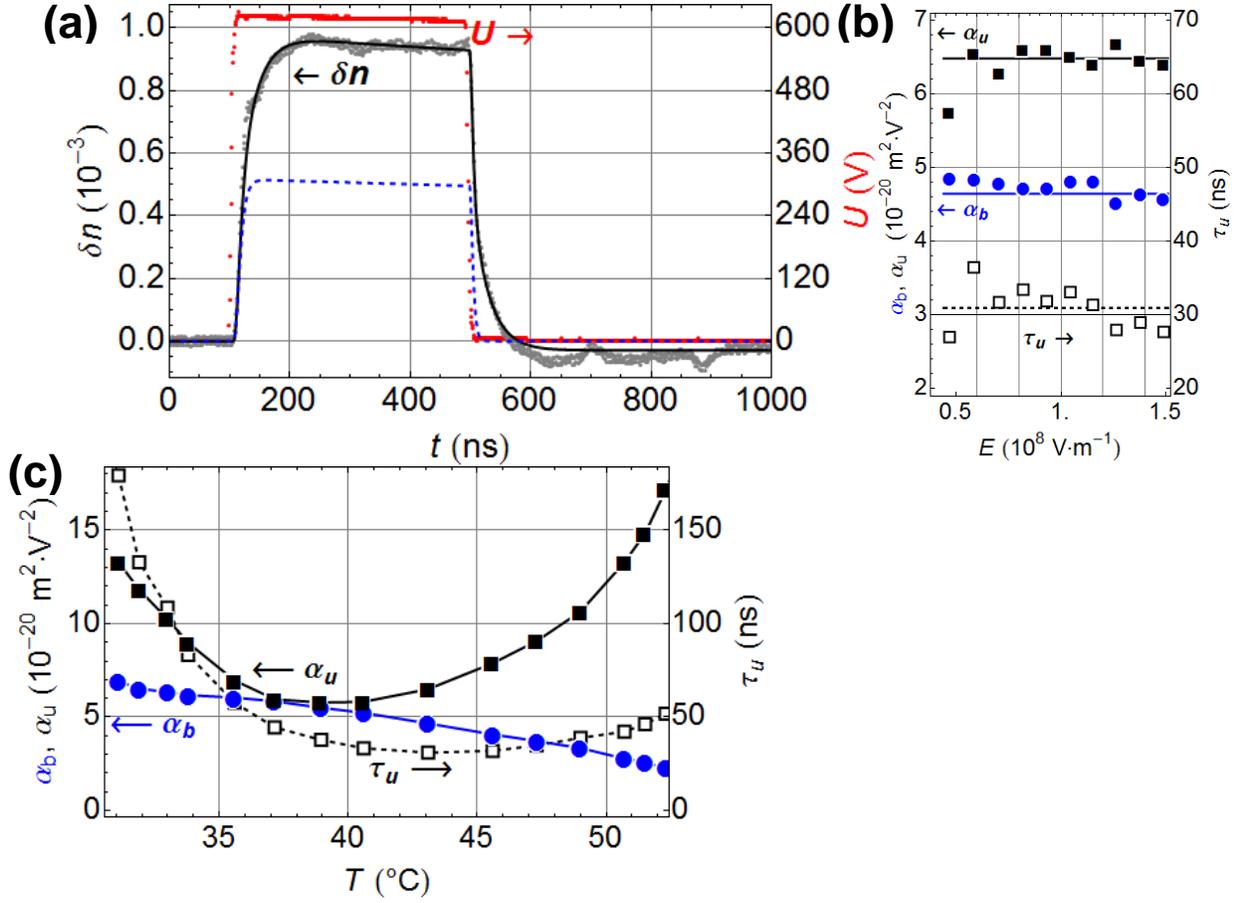

**Figure 3.** (**a**) Birefringence change (grey dots) in response to a voltage pulse (red) at $T = 43^0\text{C}$ is fitted with Eqs.(3,5), $\tau_u = 31\text{ ns}$, $\tau_b = 1.76\text{ ns}$, $\alpha_u = 6.5 \times 10^{-20}\text{ m}^2\text{ V}^{-2}$, $\alpha_b = 4.7 \times 10^{-20}\text{ m}^2\text{ V}^{-2}$ (solid black line). The blue dashed line shows the biaxial contribution with $\tau_b = 1.76\text{ ns}$ and $\alpha_b = 4.7 \times 10^{-20}\text{ m}^2\text{ V}^{-2}$. Fitting of the voltage profile by Eq. (7) with $U_0 = 626\text{ V}$, $\tau_{on} = 3.2\text{ ns}$, $\tau_{off} = 3.2\text{ ns}$, $\tau_a = 18\,\mu\text{s}$, $t_{on} = 93\text{ ns}$, and $t_{off} = 487\text{ ns}$ provides no visible deviations from experimental points. (**b**) The fitting parameters for uniaxial and biaxial contributions are field independent, as in the theoretical model. (**c**) Temperature dependences of $\alpha_u$, $\alpha_b$, and $\tau_u$.

To summarize, we describe a new electro-optic effect in liquid crystals. We call it an EMOP effect, to stress that the phenomenon is caused by electrically induced modification of the order parameter (optic tensor) rather than by realignment of the director. The main distinctive feature of this effect is an extremely fast, nanoseconds and tens of nanoseconds, response to the electric field. The fast response characterizes both the field-on and field-off stages of switching. The nanosecond dynamics of the EMOP effect are drastically different from the slow (milliseconds) field-off director relaxation in the Frederiks effect. The field-induced changes of birefringence $\delta n \sim 10^{-3}$ might appear modest. Three comments are in order here. First, the quantity of interest in electro-optical applications is focused on the phase retardation $\Gamma(t) \propto L$, where $L$ is the pathway of light that can be made large. Second, the measured values $\delta n \sim 10^{-3}$ are obtained for a material in which the natural field-free $\Delta n$ is rather small (0.03); $\Delta n$ and thus $\delta n$ can be increased through chemical synthesis. Third, the EMOP effect can be enhanced in materials that are predisposed to a local biaxial order, such as bent-core mesogens. Further improvement in the switching speed can be achieved by using ultra-short voltage pulses to trigger only the fast (biaxial) EMOP response.

EMOP is a new member of the broad family of the Kerr-type effects, in which the field induced changes of birefringence grow with the square of an applied electric field. The classic Kerr effect is observed in isotropic fluids. Past research has revealed that the Kerr effect is strongly enhanced in the vicinity of the isotropic-to-nematic phase transition [5-7, 18-21]. By staging the field-induced changes in the standard nematic and by eliminating the detrimental influence of director fluctuations, we demonstrate record-breaking response times and identify the mechanisms behind the effect, associated with uniaxial and biaxial modifications of the optic tensor. A complete description of the biaxial nematic phase requires four OPs which define the uniaxial and biaxial orders of the long and short molecular axes [22]. Our experiments are well fitted with only two relaxation processes. The reasons might be as follows. First, the relaxation times of some of the four OPs might be similar to each other. Second, the EMOP effect is likely to be affected most strongly by the following two OPs: (i) uniaxial ordering of long molecular axes and (ii) biaxial ordering of short molecular axes. These two OPs are predicted to be dominant in the spontaneous biaxial nematic phase [23, 24]. The same OPs are expected to be dominant in our experiment, since changes in the uniaxial ordering (i) cause strong changes in optical anisotropy and the biaxial ordering (ii) is affected by interactions of transverse molecular dipoles with the applied electric field. Further studies are needed to clarify the possible contributions of other OPs; this work is in progress.

The essence of EMOP embraces a much broader range of materials and driving forces than is presented in this work. Instead of the voltage pulses, one can employ the electric field of a light wave to modify the optic tensor, similar to what was extensively studied in the isotropic phase (see e.g. [25] and references therein); instead of the low-molecular weight uniaxial nematic, one can use smectics and chiral nematics, polymers, etc. The nanosecond switching of optical properties of liquid crystals opens a new chapter in the fundamental science and applications of these materials.

**Acknowledgments:** The work was supported by DOE Grant No. DE-FG02-06ER 46331. We thank P. Bos, A. Jákli, P. Palffy-Muhoray, J.V. Selinger, and D.-K. Yang for reading the manuscript and useful discussions, R. Eidenschink (Nematel GmbH) for providing us with CCN-47, and M. Groom for help with the experimental set-up.

**References:**

1. D.-K. Yang, S.-T. Wu, *Fundamentals of liquid crystal devices*. (John Wiley, Chichester ; Hoboken, NJ, 2006).

2. H. Takanashi, J. E. MacLennan, N. A. Clark, *Jpn. J. Appl. Phys.* **37**, 2587 (1998).

3. L. M. Blinov, V. G. Chigrinov, *Electrooptic effects in liquid crystal materials*. (Springer-Verlag, New York, 1994).

4. P. G. de Gennes, J. Prost, *The physics of liquid crystals*. (Clarendon Press; Oxford Univ. Press, Oxford New York, ed. 2nd, 1993).

5. C. P. Fan, M. J. Stephen, *Phys. Rev. Lett.* **25**, 500 (1970).

6. P. Palffy-Muhoray, D. A. Dunmur, *Mol. Cryst. Liq. Cryst.* **97**, 337 (1983).

7. I. Lelidis, G. Durand, *Phys. Rev. E* **48**, 3822 (1993).

8. J. A. Olivares, S. Stojadinovic, T. Dingemans, S. Sprunt, A. Jákli, *Phys. Rev. E* **68**, 041704 (2003).

9. M. Nagaraj *et al.*, *Appl. Phys. Lett.* **96**, 011106 (2010).

10. R. Stannarius, A. Eremin, M. G. Tamba, G. Pelzl, W. Weissflog, *Phys. Rev. E* **76**, 061704 (2007).

11. V. Borshch, S. V. Shiyanovskii, O. D. Lavrentovich, *Mol. Cryst. Liq. Cryst.* **559**, 97 (2012).

12. P. G. de Gennes, *C.R. Acad. Sci. Paris* **266B**, 15 (1968).

13. J. L. Martinand, G. Durand, *Solid State Commun.* **10**, 815 (1972).

14. Y. Poggi, J. C. Filippini, *Phys. Rev. Lett.* **39**, 150 (1977).

15. D. A. Dunmur, P. Palffy-Muhoray, *J. Phys. Chem.* **92**, 1406 (1988).

16. See Supplemental Material for details of adjustment of phase retardation by the Soleil-Babinet compensator.

17. L. D. Landau, I.M. Khalatnikov, *Dokl. Akad. Nauk SSSR* **46**, 469 (1954).

18. J. C. Filippini, Y. Poggi, *J. Phys. D: Appl. Phys.* **8**, L152 (1975).

19. H. J. Coles, B. R. Jennings, *Mol. Phys.* **36**, 1661 (1978).

20. S. Dhara, N. V. Madhusudana, *Eur. Phys. J. E* **22**, 139 (2007).

21. M. Gu, S. V. Shiyanovskii, O. D. Lavrentovich, *Phys. Rev. E* **78**, 040702(R) (2008).

22. J. P. Straley, *Phys. Rev. A* **10**, 1881 (1974).

23. R. Berardi, L. Muccioli, S. Orlandi, M. Ricci and C. Zannoni, *J. Phys.: Condens. Matter* **20**, 463101 (2008).

24. S. V. Shiyanovskii, *Phys. Rev. E* **87**, 060502(R) (2013).

25. Hu Cang, J. Li, V.N. Novikov, and M.D. Fayer , *J Chem. Phys*. **118**, 9303 (2003).

# Nanosecond electro-optic switching of a liquid crystal

Volodymyr Borshch, Sergij V. Shiyanovskii, and Oleg D. Lavrentovich

## Supplemental Material

**Adjustment of phase retardation by the Soleil-Babinet compensator**

An ideal response of the electro-optic effect would be described by the equation

$$I(t) = I_0 \sin^2\left\{\frac{\pi\left[\delta n(t) + \Delta n_{eff}\right]d}{\lambda} + \frac{\phi_{SB}}{2}\right\}, \tag{S1}$$

where $I_0$ is the intensity of light impinging on the NLC cell. By changing the phase retardation $\phi_{SB}$ of the Soleil-Babinet compensator, one can change the intensity of a measured light from its absolute minimum, $I = 0$, to an absolute maximum $I = I_0$. In reality, the experimentally measured signal is of the form

$$I(t) = \left[I_{max}(t) - I_{min}(t)\right]\sin^2\left\{\frac{\pi\left[\delta n(t) + \Delta n_{eff}\right]d}{\lambda} + \frac{\phi_{SB}}{2}\right\} + I_{min}(t), \tag{S2}$$

in which the minimum value $I_{min}$ is different from 0 (say, because of the incoherent light leakage) and $I_{max}$ is different from $I_0$ (for example, because of light reflection at interfaces, light scattering and absorption in all the elements of the system, including the NLC slab). The time dependence of both $I_{max}$ and $I_{min}$ reflects the fact that these quantities might depend on the presence of the electric pulse. In order to eliminate the effects of non-ideality and to extract the values of $\delta n(t)$, we performed the light intensity measurements with two settings of the Soleil-Babinet phase difference $\phi_{SB}$.

The Soleil-Babinet compensator is set in two positions, with two different values of phase retardation, $\phi_A = \frac{2\pi}{\lambda}\left(\frac{\lambda}{4} - \Delta n_{eff} d\right)$ and $\phi_B = \frac{2\pi}{\lambda}\left(\frac{3\lambda}{4} - \Delta n_{eff} d\right)$, that correspond to the same transmitted intensity in the field-free state, $I_{A,B}(t=0) = \left[I_{max}(0) + I_{min}(0)\right]/2$, see Fig. S1a.

The change $\Delta I_{A,B}(t) = I(t, \phi_{A,B}) - I(0, \phi_{A,B})$ of intensity in these points are then controlled by changes of birefringence $\delta n(t)$ and intensities $\Delta I_{max}(t) = I_{max}(t) - I_{max}(0)$ and $\Delta I_{min}(t) = I_{min}(t) - I_{min}(0)$. These changes are small, thus we neglect the second and higher order terms:

$$\Delta I_{A,B}(t) = \pm\left[I_{max}(0) - I_{min}(0)\right]\frac{\pi \delta n(t) d}{\lambda} + \frac{\Delta I_{max}(t) + \Delta I_{min}(t)}{2}, \tag{S3}$$

where "+" and "−" correspond to A and B points, respectively. Measuring in these two points provides a two-fold benefit: (1) it maximizes the sensitivity of measurement, and (2) it separates the field-induced birefringence $\delta n(t)$ effect of interest from the parasitic effects associated with $\Delta I_{max}$ and $\Delta I_{min}$. These two contributions can be discriminated from the "ideal response" by evaluating the half-sum $\Delta I_+(t) = \left[\Delta I_A(t) + \Delta I_B(t)\right]/2$ and the half-difference $\Delta I_-(t) = \left[\Delta I_A(t) - \Delta I_B(t)\right]/2$ of the optical measurements recorded in the points $\phi_{SB} = \phi_A$ and $\phi_{SB} = \phi_B$ (Fig.S1a and S1c). For the data analysis, we used the half-difference $\Delta I_-(t)$, which corresponds to the "ideal response" of the birefringence effects $\delta n(t)$. As seen in Fig.S1c, the half-difference $\Delta I_-(t)$ signal is significantly larger than the half-sum $\Delta I_+(t)$ signal, which indicates the prevalence of the field-induced birefringence $\delta n(t)$ effect over the parasitic factors.

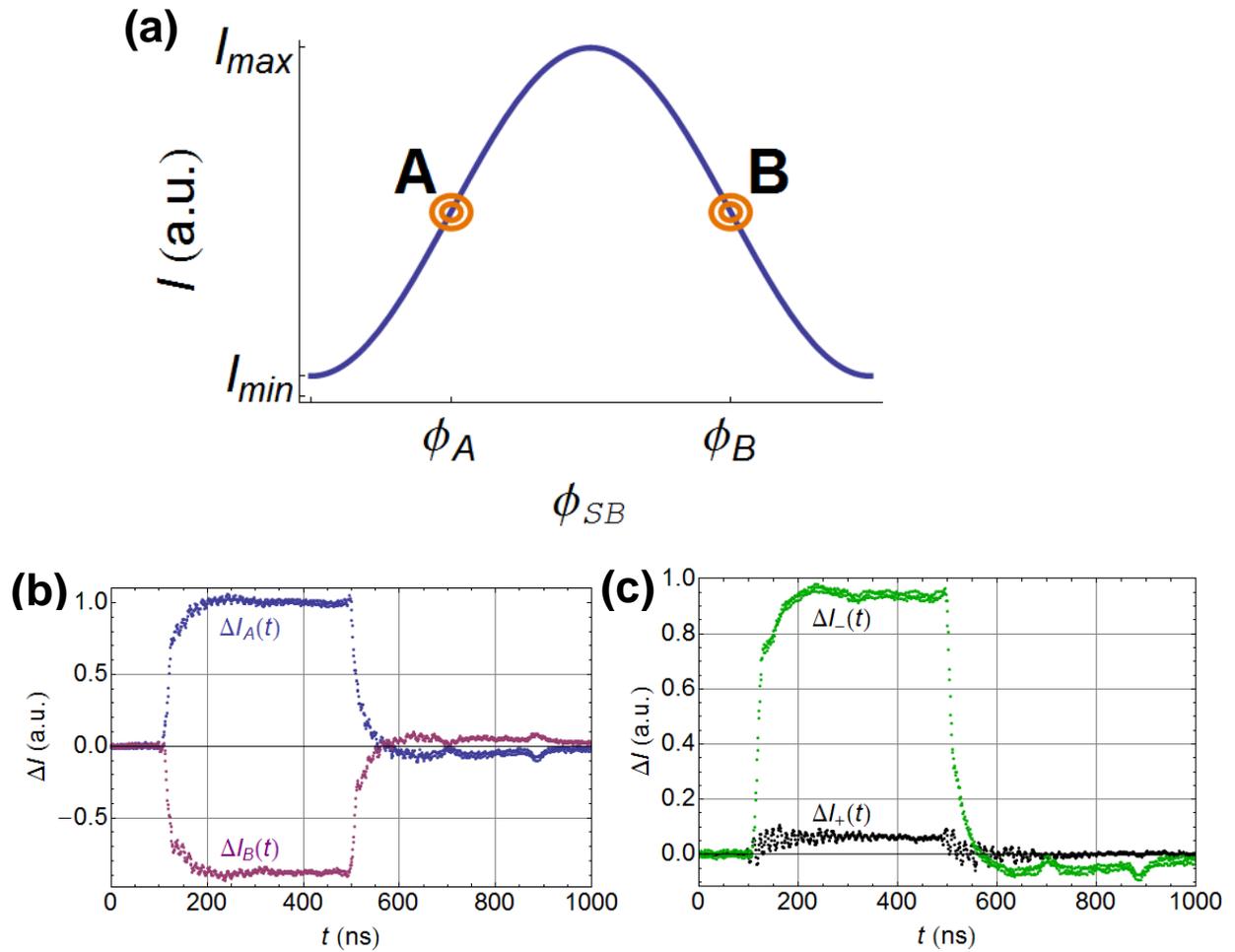

**Fig. S1.** (**a**) Two positions of the Soleil-Babinet compensator allow working in the linear area of the intensity curve where the small change of the optical phase retardation corresponds to the maximum change of the intensity. The positive and negative slopes allow separating the birefringence and light scattering effects. (**b**) The optical response measured to $U_0 = 626$ V pulse at $T = 43$ °C. The blue line was measured with $\phi_{SB} = \phi_A$ and the purple line with $\phi_{SB} = \phi_B$ position of the Soleil-Babinet compensator. (**c**) The green line shows the half-difference $\Delta I_-(t)$, and the black line shows the half-sum $\Delta I_+(t)$ of the two optical response curves shown in Fig.S1b.